\documentclass[prd,twocolumn,groupedaddress,showpacs]{revtex4}

\begin{document}
% You should use BibTeX and revtex.bst for references
\bibliographystyle{apsrev}

\preprint{COLO-HEP-445}
\preprint{May 2000}

\title{From CKM Matrix to MNS Matrix: A Model Based on
Supersymmetric $SO(10) \times U(2)_{F}$ Symmetry} 

% Optional argument for running titles on pages
%\title[]{}


% \affiliation command. The \affiliation command MUST follow the

\author{Mu-Chun Chen}
\email[]{mu-chun.chen@colorado.edu}
%\thanks{}
\affiliation{Department of Physics\\
             University of Colorado\\
             Boulder, CO 80309-0390}

\author{K.T. Mahanthappa}
\email[]{ktm@verb.colorado.edu}
%\thanks{}
\affiliation{Department of Physics\\
             University of Colorado\\
             Boulder, CO 80309-0390}
%\date{\today}

\begin{abstract}
We construct a realistic model based on SUSY $SO(10)$ with $U(2)$ flavor
symmetry. In contrast to the commonly used effective operator approach, 
$126-$dimensional Higgses are used to construct the Yukawa sector. R-parity symmetry
is thus preserved at low energies. The Dirac and right-handed Majorana mass
matrices in our model have very small mixing, and they combine with the seesaw
mechanism resulting in a large leptonic mixing. The symmetric mass textures
arising from the left-right symmetry breaking chain of $SO(10)$
give rise to very good predictions; 15 masses (including 3 right-handed
Majorana neutrino masses) and 6 mixing angles are predicted by 11
parameters. Both the vacuum oscillation and LOW solutions are favored for 
the solar neutrino problem.
\end{abstract}



%\maketitle must follow title, authors, abstract and PACS
\maketitle

%\section{Introduction}
%\label{intro}
%\subsection{}
%\subsubsection{}

The flavor problem with hierarchical fermion masses and mixing has attracted a
great deal of attention especially since the advent of the atmospheric
neutrino oscillation data from Super-Kamiokande \cite{SuperK:1998a} indicating
non-zero neutrino masses. The non-zero neutrino masses give support to the
idea of grand unification based on $SO(10)$ in which all the 16 fermions
(including the right-handed neutrinos) can be accommodated in one single spinor
representation. Furthermore, it provides a framework in which seesaw
mechanism arises naturally. Models based on $SO(10)$ (and some with $E_{6}$)
combined with a continuous or discrete flavor symmetry group have been
constructed to understand the flavor problem. Most of the recent ones have
used asymmetric or "lopsided" mass textures to account for the maximal mixing
in the neutrino sector. Symmetric mass textures have less parameters and hence
could lead to more predictive power. Naively one expects, for symmetric mass 
textures, six texture zeros in the quark sector. But it has been observed by
Ramond, Roberts and Ross \cite{Ramond:fmass1993a} that the highest number 
of texture zeros has to be five, and 
using phenomenological analyses, they were able to 
arrive at five sets of up- and down-quark mass matrices with five texture zeros. 
Our analysis with recent experimental data and using CP conserving real symmetric 
matrices indicates that only one set (labeled set (v) in \cite{Ramond:fmass1993a}) remains 
viable (see below). 
The aim of this paper is to construct a
realistic model based on $SO(10)$ combined with $U(2)$ as the flavor group, 
utilizing this set of symmetric mass textures for charged
fermions.
We first discuss the viable phenomenology of
mass textures followed by the model which accounts for it, and then the 
implications of the model for neutrino mixing are presented.

%\section{Mass texture analysis} %\label{sec:texture}
%\subsection{}
%\subsubsection{}

{\bf Mass Texture Analysis:}
Throughout this paper we consider CP conserving real mass matrices. We do not
lose any generality in our results since the CP violating phases do not have
significant contributions to other parameters. A more detailed analysis taking
into account CP violating phases will be given
elsewhere \cite{chen:fmass2000a}. 

We consider the following mass textures at the GUT scale for the up-quark,
down-quark and charged lepton sectors \cite{Ramond:fmass1993a}, 
%\begin{widetext}
\begin{eqnarray}
\label{eq:texture}
M_{u} =
\left( \begin{array}{ccc}
0 & 0 & a \\
0 & b & c \\
a & c & 1
\end{array} \right) d, 
\qquad &
M_{d}=
\left( \begin{array}{ccc}
0 & e & 0 \\
e & f & 0 \\
0 & 0 & 1
\end{array} \right) h 
\qquad
\nonumber\\
\vspace{1cm}
M_{e}= 
\left( \begin{array}{ccc}
0 & e & 0 \\
e & -3f & 0 \\
0 & 0 & 1
\end{array} \right) h &
\end{eqnarray}
%\end{widetext}
with $a \simeq b \ll c \ll 1$ and $e \ll f \ll 1$.
After diagonalizing $M^{\dag}M$, one
obtains the following non-negative mass eigenvalues and mass ratios:
%\begin{widetext}
\begin{eqnarray}
\label{eq:masses}
m_{u} \simeq \frac{a^{2} b d}{-b + c^{2}},&\quad
m_{c} \simeq (-b+c^{2}) d,&\quad
m_{t} \simeq d \nonumber\\
m_{d} \simeq \frac{e^{2} f h}{e^{2}+f^{2}},&\quad
m_{s} \simeq \frac{(2e^{2} f + f^{3}) h}{e^{2} + f^{2}},&\quad
m_{b} = h \nonumber\\
m_{e} \simeq \frac{3e^{2} f h}{e^{2} + 9 f^{2}},&\qquad
m_{\mu} \simeq \frac{6e^{2} f h + 27 f^{3} h}{e^{2} + 9 f^{2}},&\quad
m_{\tau} = h
\end{eqnarray}
%\end{widetext}
%\begin{widetext}
\begin{eqnarray}
\frac{m_{d}}{m_{e}} \simeq \frac{e^{2}+9f^{2}}{3(e^{2}+f^{2})} 
\simeq 3 + O(\frac{e^{2}}{f^{2}}) 
\nonumber\\
\frac{m_{s}}{m_{\mu}} \simeq \frac{2e^{4}+19e^{2}f^{2}+9f^{4}}{6e^{4}
+33e^{2}f^{2}+27f^{4}} \simeq \frac{1}{3} + O(\frac{e^{2}}{f^{2}}) 
\nonumber\\
\frac{m_{b}}{m_{\tau}} = 1
\end{eqnarray}
%\end{widetext}
These analytic expressions are very good approximations to the exact
eigenvalues. It can be easily seen that the phenomenologically favored
Georgi-Jarlskog relations \cite{Georgi:fmass1979a}
%%,Georgi:fmass1979b} 
are obtained
%\begin{widetext} 
\begin{equation} 
\label{eq:GJ}
m_{d} \simeq 3 m_{e},
\qquad
m_{s} \simeq \frac{1}{3} m_{\mu},
\qquad
m_{b} = m_{\tau}
\end{equation}
%\end{widetext}
As we will see later, these relations between the
down-quark sector and the charged lepton sector can be naturally achieved
in $SO(10)$. 

In order to explain the smallness of the neutrino masses, we will adopt
the type I seesaw mechanism \cite{Gell-Mann:fmass1979a} which requires both
Dirac and right-handed Majorana mass matrices to be present in the
Lagrangian. The right-handed Majorana mass matrix $M_{\nu_{RR}}$ is at present
an unknown sector. The only constraint is that it must be constructed in such
a way that it gives a favored low energy Majorana neutrino mass
matrix $M_{\nu_{LL}}$ via the seesaw mechanism. We first consider the low
energy (left-handed) Majorana neutrino mass matrix to get some insights into
the structure of $M_{\nu_{RR}}$. 
We adopt the hierarchical scenario: $|m_{\nu_{3}}| \gg |m_{\nu_{2}}|,
|m_{\nu_{1}}|$, to accommodate the experimental neutrino oscillation data.  One
way to achieve a large mixing in the $\nu_{\mu}-\nu_{\tau}$ sector and at the
same time a large mass splitting between $m_{\nu_{2}}$ and $m_{\nu_{3}}$
is to consider
\begin{equation}   
\label{eq:Mll}  
M_{\nu_{LL}} \sim \left( 
\begin{array}{ccc}  
0 & 0 & t \\
0 & 1 & 1 \\  
t & 1 & 1 
\end{array} \right) \Lambda
\end{equation}
A generic feature of the mass matrix of this type is that it leads to a
large mixing in both $\nu_{e}-\nu_{\mu}$ and $\nu_{\mu}-\nu_{\tau}$ sectors 
(the so-called bimaximal mixing) for a broad range of $t$, $0 \leq t \leq 1$. 
In order to have a large mass splitting, we require $t \ll 1$. The three
eigenvalues of this mass matrix keeping only the dominant orders are given by,
in units of $\Lambda$,
%\begin{widetext}
\begin{eqnarray}
|m_{1}| & \simeq & 
\frac{t}{\sqrt{2}} - \frac{t^{2}}{8} - \frac{3t^{3}}{64\sqrt{2}} \nonumber\\
|m_{2}| & \simeq &
\frac{t}{\sqrt{2}} + \frac{t^{2}}{8} - \frac{3t^{3}}{64\sqrt{2}} \nonumber\\
|m_{3}| & \simeq & 
2 + \frac{t^{2}}{4}
\end{eqnarray}
%\end{widetext}
The diagonalization matrix up to order $O(t^{2})$ is given by
\begin{widetext}
\noindent
\rule{8.8cm}{.3pt}%
\rule{.3pt}{.25cm}  
\begin{equation}
U_{\nu_{LL}} = 
\left( \begin{array}{ccc}
\frac{1}{\sqrt{2}} - \frac{t}{16} - \frac{17 t^{2}}{256 \sqrt{2}}  &
-\frac{1}{2} - \frac{5t}{16 \sqrt{2}} - \frac{23 t^{2}}{512}  &
\frac{1}{2} - \frac{3t}{16 \sqrt{2}} - \frac{25 t^{2}}{512}  \\
-\frac{1}{\sqrt{2}} - \frac{t}{16} + \frac{17 t^{2}}{256 \sqrt{2}}  &
-\frac{1}{2} + \frac{5t}{16 \sqrt{2}} - \frac{23 t^{2}}{512}  &
\frac{1}{2} + \frac{3t}{16 \sqrt{2}} - \frac{25 t^{2}}{512}  \\
\frac{t}{2 \sqrt{2}}&
\frac{1}{\sqrt{2}} - \frac{3 t^{2}}{16 \sqrt{2}}  &
\frac{1}{\sqrt{2}} + \frac{3 t^{2}}{16 \sqrt{2}}
\end{array} \right)
\end{equation}
\end{widetext}
Note that what the neutrino mixing matrix really means is the mismatch between
the charged lepton flavor basis and the neutrino flavor basis analogous to
the Cabbibo-Kobayashi-Maskawa quark mixing matrix $V_{CKM}$,
and is the Maki-Nakagawa-Sakata matrix $U_{MNS}$ defined as, 
%\begin{widetext}
\begin{eqnarray}
U_{MNS}  \equiv  U_{e_{L}}U_{\nu_{LL}}^{\dag} \nonumber\\
=  \left(\begin{array}{ccc}
U_{e\nu_{1}} & U_{e\nu_{2}} & U_{e\nu_{3}}\\
U_{\mu\nu_{1}} & U_{\mu\nu_{2}} & U_{\mu\nu_{3}}\\
U_{\tau\nu_{1}} & U_{\tau\nu_{2}} & U_{\tau\nu_{3}}\\
\end{array} \right)
\end{eqnarray}  
%\end{widetext} 
Since the mixing matrix in the charged lepton sector $U_{e_{L}}$ is
almost diagonal, combining $U_{\nu_{LL}}^{\dag}$ and $U_{e_{L}}$ results in a
nearly bimaximal mixing pattern in the lepton mixing matrix $U_{MNS}$.
The squared mass difference between $m_{\nu_{1}}^{2}$ and
$m_{\nu_{2}}^{2}$ is of the order of $O(t^{3})$ while the squared mass
difference between $m_{\nu_{2}}^{2}$ and $m_{\nu_{3}}^{2}$ is of the
order $O(1)$. It is clear that the mass matrix eq.(\ref{eq:Mll}) naturally
leads to the phenomenologically favored result 
\begin{equation}
\vert\Delta m_{23}^{2}\vert \gg \vert\Delta m_{12}^{2}\vert
\end{equation}
Depending on the value of $t$, both vacuum oscillation (VO) solution and
large angle MSW (LAMSW) solution are possible. The VO solution suggests that 
$\frac{\Delta m_{\odot}^{2}}{\Delta m_{atm}^{2}} \simeq 10^{-7}$. Since  
$\frac{\Delta m_{\odot}^{2}}{\Delta m_{atm}^{2}} \sim t^{3}$, one can see
immediately that $t \sim 10^{-3}$. On the other hand, the LAMSW
solution suggests that 
$\frac{\Delta m_{\odot}^{2}}{\Delta m_{atm}^{2}} \simeq 10^{-2}$, and $t$ is
then required to be $\sim 10^{-1}$. Since the
element $U_{e\nu_{3}}$ in $U_{MNS}$  is proportional to $t$, 
an accurate  measurement of $U_{e\nu_{3}}$ thus could provide some hints to
single out one of the solar oscillation solutions if the neutrino mixing
pattern is indeed bimaximal.

We assume that the Dirac neutrino mass matrix has the same texture (that
is, positions of the zeros) as the up-quark mass matrix 
\begin{equation} 
\label{eq:Mnu}
M_{\nu_{LR}}=
\left( \begin{array}{ccc}
0 & 0 & \alpha \\
0 & \beta & \gamma \\
\alpha & \gamma & 1
\end{array} \right) \eta
\end{equation}
with $\alpha \simeq \beta \ll \gamma \ll 1 $.
We see later that $M_{u}$ and $M_{\nu_{LR}}$ can be in
fact identical in $SO(10)$. To achieve
$M_{\nu_{LL}}$ of the form of eq.(\ref{eq:Mll}) one needs a right-handed
neutrino Majorana mass matrix of the same texture as $M_{\nu_{LR}}$
\begin{equation} 
\label{eq:Mrr}
M_{\nu_{RR}}=
\left( \begin{array}{ccc}
0 & 0 & \delta_{1} \\
0 & \delta_{2} & \delta_{3} \\
\delta_{1} & \delta_{3} & 1
\end{array} \right) M_{R}
\end{equation}
with
\begin{displaymath}
\delta_{1}, \delta_{2}, \delta_{3} \ll 1.
\end{displaymath}
%\begin{widetext}
\begin{eqnarray}
\label{eq:delta}
\delta_{1} \simeq \frac{\alpha^{2}}{2\alpha-2\alpha\gamma+\gamma^{2}t},\quad &
\delta_{2} \simeq
\frac{\beta^{2}t}{2\alpha-2\alpha\gamma+\gamma^{2}t}\nonumber\\ 
\delta_{3}
\simeq \frac{\alpha(\gamma-\beta)+\beta\gamma t}
{2\alpha-2\alpha\gamma+\gamma^{2}t}&
\end{eqnarray}
%\end{widetext}
After seesaw mechanism takes place,
\begin{equation}
\label{eq:seesaw}
M_{\nu_{LL}} = -M_{\nu_{LR}}^{T} M_{\nu_{RR}}^{-1} M_{\nu_{LR}}
\end{equation}
$M_{\nu_{LL}}$ of the form eq.(\ref{eq:Mll}) results. It is interesting to see
that the matrix operation in eq.(\ref{eq:seesaw}) is form invariant. That is to
say, $M_{\nu_{LL}}$ has the same texture as that of $M_{\nu_{LR}}$ and
$M_{\nu_{RR}}$. Since $\delta_{1}, \delta_{2}$ and $\delta_{3}$ are much
smaller than $1$, the mixing in the right-handed neutrino mass matrix
$M_{\nu_{RR}}$ is generally small. Since the mixings in both
$M_{\nu_{LR}}$ and $M_{\nu_{RR}}$ are small, our model falls into the
category that the large neutrino mixing is purely due to
the matrix operations in the seesaw mechanism given that charged lepton, Dirac
neutrino and right-handed neutrino mixings are small, as classified in
Ref.\cite{Barr:fmass2000a}.
We note that with the structure of $M_{\nu_{LR}}$ in eq.(\ref{eq:Mnu}) we find
it hard, though not impossible, to accommodate the small angle MSW solution in
our model \cite{chen:fmass2000a}.

We emphasize that, in the neutrino sector, we have been able to get a large
mixing and at the same time a large mass splitting by using a symmetric Dirac
mass matrix and a hierarchical right-handed mass matrix with very small
mixing; the latter gives three superheavy hierarchical right-handed
neutrino masses. Asymmetric Dirac mass matrices have been used before to get a
large mixing and a large mass splitting
 \cite{Altarelli:1998nx,
%%,Altarelli:1998ns,Altarelli:1998sr,Altarelli:1999gu,
%%Altarelli:1999dg,Altarelli:1999wi,
Babu:1998wi,
Albright:1998vf,Berezhiani:fmass1999a,U1A}.
%%,Albright:1998sy,Albright:fmass2000a,Albright:fmass2000b}.

%\section{$U(2)$ as a flavor symmetry}
%\label{sec:framework}
%\subsection{}
%\subsubsection{}

{\bf $\mathbf{U(2)}$ as a Flavor Symmetry:}
A prototype scenario which produces hierarchy in the fermion mass matrices is 
the Froggatt-Nielsen mechanism \cite{Froggatt:fmass1979a}.
It simply says that the heaviest matter fields acquire their masses through
tree level interactions with the Higgs fields while masses of lighter matter
fields are produced by higher dimensional interactions involving, in addition
to the regular Higgs fields, exotic vector-like pairs of matter fields and the
so-called flavons (flavor Higgs fields). After integrating out superheavy $(
\approx M)$ vector-like matter fields, the mass terms of the light matter
fields get suppressed by a factor of $\frac{<\theta>}{M}$, where $<\theta>$ is
the VEVs of the flavons and $M$ is the UV-cutoff of the effective theory above
which the flavor symmetry is exact. We assume $M \gg M_{GUT}$. We choose
$U(2)$ as the flavor symmetry group
 \cite{Barbieri:1996uv}
%%,Barbieri:1997ae,
%%Barbieri:fmass1997a,Carone:fmass1997a,Romanino:1997an,Barbieri:fmass1999a,
%%Barbieri:fmass1999b,Hall:fmass1999a}. 
which has two attractive
features: (i) it gives rise to the degeneracies between 1-2 families needed to
suppress the supersymmetric FCNC in the squark sector, and (ii) a multi-step
breaking of $U(2)$ gives rise to the observed inter-family hierarchy
naturally. Unlike models based on the most commonly used $U(1)$ symmetry, in
which one has the freedom in choosing $U(1)$ charges for various matter
fields, a $U(2)$ flavor symmetry appears to be a much more constrained
framework for constructing realistic models. The basic idea is very simple.
The three families of matter fields transform under a $U(2)$ flavor symmetry
as 
\begin{equation}  \psi_{a} \oplus \psi_{3} = 2 \oplus 1
\end{equation}
where $a =1,2$ and the subscripts refer to family indices. In the symmetric
limit, only the third family of matter fields have non-vanishing Yukawa
couplings. This can be understood easily since the third family of matter
fields have much higher masses compared to the other two families of matter
fields. $U(2)$ breaks down in two steps:
\begin{equation}
\label{eq:steps} U(2) \stackrel{\epsilon M}{\longrightarrow} 
U(1) \stackrel{\epsilon' M}{\longrightarrow}
nothing
\end{equation}
with $\epsilon' \ll \epsilon \ll 1$ and $M$ is the UV cut-off of the 
effective theory mentioned before. These small
parameters $\epsilon$ and $\epsilon'$ are the ratios of the vacuum expectation
values of the flavon fields to the cut-off scale. Note that since 
\begin{eqnarray} 
\psi_{3}\psi_{3} \sim 1_{S},\qquad \psi_{3}\psi_{a} \sim 2\nonumber\\ 
\psi_{a}\psi_{b} \sim 2 \otimes 2 = 1_{A} \oplus 3 
\end{eqnarray}
the only relevant flavon fields are in the $1_{A}, 2$ and $3$
dimensional representations of $U(2)$, namely, 
\begin{equation}
A^{ab} \sim 1_{A}, \qquad \phi^{a} \sim 2, \qquad S^{ab} \sim 3
\end{equation}
Because we are confining ourselves to symmetric mass textures, we use only
$\phi^{a}$ and $S^{ab}$. 
Since all the $16$ observed matter fields of each family fall
nicely into a $16-$dimensional spinor representation of $SO(10)$, the most
general superpotential that generates fermion masses for a $SO(10) \times U(2)$
model has the following very simple form 
\begin{equation} 
W = H(\psi_{3}
\psi_{3} + \psi_{3} \frac{\phi^{a}}{M} \psi_{a} + \psi_{a}\frac{S^{ab}}{M}
\psi_{b}) 
\end{equation}
In a specific $U(2)$ basis,
%\begin{widetext}
\begin{equation}
\label{genU2}
\frac{ \left< \phi \right> }{M} \sim O \left( \begin{array}{c}
\epsilon' \\
\epsilon
\end{array} \right),
\qquad
\frac{ \left< S^{ab} \right> }{M} \sim O \left( \begin{array}{cc}
\epsilon' & \epsilon' \\
\epsilon' & \epsilon
\end{array} \right)
\end{equation}
%\end{widetext}
Here we have indicated the VEVs \textit{all} the flavon fields could acquire
for symmetry breaking in eq.(\ref{eq:steps}). 
The mass matrix would take the following form
\begin{equation}
M \sim O \left( \begin{array}{ccc}
\epsilon' & \epsilon' & \epsilon' \\
\epsilon' & \epsilon & \epsilon \\
\epsilon' & \epsilon & 1 
\end{array} \right)
\end{equation}

In $SO(10)$, at the renormalizable level, only three types of Higgs fields can
couple to fermions,
\begin{equation}
16 \otimes 16 = 10_{S} \oplus 120_{A} \oplus 126_{S}
\end{equation}
namely, $10$, $120_{A}$, and $\overline{126}_{S}$, where the subscripts $S$ and $A$
refer to the symmetry property under interchanging two family indices in
the Yukawa couplings $\mathcal{Y}_{ab}$. That is,
%\begin{widetext}
\begin{equation}
\mathcal{Y}_{ab}^{10} = \mathcal{Y}_{ba}^{10},\quad
\mathcal{Y}_{ab}^{120} = -\mathcal{Y}_{ba}^{120},\quad
\mathcal{Y}_{ab}^{\overline{126}} = \mathcal{Y}_{ba}^{\overline{126}}
\end{equation}
%\end{widetext}
$\phi^{a}$ and $S^{ab}$ can couple to only $10_{S}$ and $\overline{126}_{S}$;
$120_{A}$ has no role in giving rise to mass textures. Note that $SO(10)$ can
break down to SM through many different breaking chains.
\
Different breaking
chains give rise to different mass relations among the up-quark, down-quark,
charged lepton and neutrino sectors. Since we are interested in symmetric mass
textures, a natural choice is the left-right symmetric route, that is,
%\begin{widetext}
\begin{equation}
\label{eq:SB}
\begin{array}{lll}
SO(10) & \longrightarrow & SU(4) \times SU(2)_{L} \times SU(2)_{R}\\
 & \longrightarrow & SU(3) \times SU(2)_{L} \times SU(2)_{R} \times
U(1)_{B-L}\\  
& \longrightarrow & SU(3) \times SU(2)_{L} \times U(1)_{Y} \\
& \longrightarrow & SU(3) \times U(1)_{EM} 
\end{array}
\end{equation}
%\end{widetext}
We have the up-quark sector related to the neutrino sector, and
the down-quark sector to the charged lepton sector. A Clebsch-Gordon
coefficient $(-3)$ appears in the lepton sectors when the $SU(4) \times
SU(2)_{L} \times SU(2)_{R}$ components $(15,2,2)$ in
$\overline{126}$ are involved in the Yukawa couplings. This factor of $(-3)$
is very crucial for obtaining the Georgi-Jarlskog relations as we have seen in
the previous section.  The general fermion Dirac mass matrices are
thus given schematically by
%\vspace{3cm} 
%\begin{widetext}
\begin{eqnarray}
M_{u}\sim\mathcal{Y}^{10}_{ab} \left< 10^{+} \right> 
+ \mathcal{Y}^{\overline{126}}_{ab} \left< \overline{126}^{+} \right>
%\end{equation}
\nonumber\\
%\begin{equation}
M_{d}\sim\mathcal{Y}^{10}_{ab} \left< 10^{-} \right> 
+ \mathcal{Y}^{\overline{126}}_{ab} \left< \overline{126}^{-} \right>
%\end{equation}
\nonumber\\
%\begin{equation}
M_{e} \sim \mathcal{Y}^{10}_{ab} \left< 10^{-} \right>
-3 \mathcal{Y}^{\overline{126}}_{ab} \left< \overline{126}^{-} \right>
%\end{equation}
\nonumber\\
%\begin{equation}
M_{\nu_{LR}} \sim \mathcal{Y}^{10}_{ab} \left< 10^{+} \right> 
-3 \mathcal{Y}^{\overline{126}}_{ab} 
\left< \overline{126}^{+}\right>
\end{eqnarray}
%\end{widetext}
and general Majorana mass matrices are given by
%\begin{widetext}
\begin{equation}
M_{\nu,RR} \sim \mathcal{Y}^{\overline{126}}_{ab} \left< \overline{126}'^{0}
\right>
\end{equation}
\begin{equation}
M_{\nu,LL} \sim \mathcal{Y}^{\overline{126}}_{ab} \left<
\overline{126}'^{+} \right> 
\end{equation}
%\end{widetext}
where various VEVs are those of the neutral components of $SO(10)$
representations as indicated below (with subscripts referring to the symmetry
groups on the r.h.s. of eq.(\ref{eq:SB}); and $+/0/-$ referring to the sign of
the hypercharge Y).
\begin{widetext}
\noindent
\rule{8.8cm}{.3pt}%
\rule{.3pt}{.25cm}     
\begin{equation}
\begin{array}{l}
\left< 10^{+} \right> : (1,0)_{31}
\subset (1,2,1)_{321} \subset 
(1,2,2,0)_{3221} \subset (1,2,2)_{422} \subset 10 \\ 
\left< 10^{-} \right> : (1,0)_{31}
\subset (1,2,-1)_{321} \subset
(1,2,2,0)_{3221} \subset (1,2,2)_{422} \subset 10
\end{array}
\end{equation}
\begin{equation}
\begin{array}{l}
\left< \overline{126}^{+} \right> : (1,0)_{31} 
\subset (1,2,1)_{321} \subset
(1,2,2,0)_{3221} \subset (15,2,2)_{422} \subset \overline{126} \\
\left< \overline{126}^{-} \right> : (1,0)_{31}
\subset (1,2,-1)_{321} \subset
(1,2,2,0)_{3221} \subset (15,2,2)_{422} \subset \overline{126} \\
\left< \overline{126}'^{0} \right> : (1,0)_{31}
\subset (1,1,0)_{321} \subset
(1,1,3,-2)_{3221} \subset (10,1,3)_{422} \subset \overline{126}\\
\left< \overline{126}'^{+} \right> : (1,0)_{31}
\subset (1,3,2)_{321} \subset
(1,3,1,2)_{3221} \subset (\overline{10},3,1)_{422} \subset \overline{126} \\
\end{array}
\end{equation}
\end{widetext}

A remark is in order here. Some models avoid the use of $\overline{126}$ dimensional
Higgses by introducing nonrenormlaizable operators of the form
$f_{a}f_{b}(16)_{h}(16)_{h}$. Such models appear to be less constrained
due to the inclusion of nonrenormalizable operators. Also, a discrete
symmetry, the R-parity symmetry, must be imposed by hand to avoid
dangerous Baryon number violating terms in the effective potential at low
energies which otherwise could lead to fast proton decay rate. Here we use
$\overline{126}$ dimensional representation of Higgses which has the advantage
that R-parity symmetry is automatic \cite{Mohapatra:rparity1986a}.
The $\overline{126}$ representation has been used in model building before
 \cite{Aulakh:1999cd}.
It is to be noted that the contribution of the $\overline{126}$-dimensional
representation to the $\beta$-function makes the model nonperturbative 
(with the onset of the Landau pole) above the unification scale $M_{GUT}$.
One could view our model as an effective theory valid below this scale
where coupling constants are perturbative.

Other breaking chains of $SO(10)$ have been considered resulting in
various interesting mass textures and thus mass relations. For
example,
 \cite{Albright:1998vf}
considers $SO(10)$ breaking through $SU(5)$ to SM and obtains
the so-called "lopsided" mass textures due to the fact that $SU(5)$ gives the
relation $M_{d}=M_{e}^{T}$. A large lepton mixing $(U_{MNS})$ arises in this
class of models from a large left-handed charged lepton mixing which relates
to a large mixing in the right-handed down-quark sector.

%\section{A model based on $SO(10) \times U(2)$}
%\label{sec:model}

{\bf A Model Based on $\mathbf{SO(10) \times U(2)}$:}
We now demonstrate how the above phenomenology emerges from
a model based on $SO(10) \times U(2)$. Here we only present the Yukawa
sector. A more complete account of the Higgs potential including symmetry
breaking sector and the doublet-triplet splitting sector will be given
elsewhere \cite{chen:fmass2000a}.  

In order to uniquely specify the Yukawa superpotential without any unwanted
interaction terms, we need to introduce $Z_{2} \times Z_{2} \times Z_{2}$
discrete symmetry. The fields needed are indicated below\\  
{\sl Matter fields:}
\begin{eqnarray} 
\psi_{a} \sim (16,2)^{-++} & \quad (a=1,2) \nonumber\\ 
\psi_{3} \sim (16,1)^{+++} & \quad
\end{eqnarray}
{\sl Higgs fields for the mass matrices:}
\begin{eqnarray}
(10,1):\quad & T_{1}^{+++},\quad T_{2}^{-+-},\quad
T_{3}^{--+}\nonumber\\
 & T_{4}^{---}, \quad T_{5}^{+--} \nonumber\\ 
(\overline{126},1):\quad & \overline{C}^{---}, \quad \overline{C}_{1}^{+++},
\quad \overline{C}_{2}^{++-}  
\end{eqnarray}
{\sl Flavon fields:}
\begin{eqnarray}
(1,2): \quad & \phi_{(1)}^{++-}, \quad \phi_{(2)}^{+-+}, \quad \Phi^{-+-}
\nonumber\\ 
(1,3): \quad & S_{(1)}^{+--}, \quad S_{(2)}^{---}, \quad
\Sigma^{++-} 
\end{eqnarray}
Note that, the entries in the parenthesis indicate the $SO(10)$ and $U(2)$
representations respectively. The superscript $+/-$ indicates the charges
under $Z_{2} \times Z_{2} \times Z_{2}$ symmetry. Various Higgs fields acquire
VEVs in the following directions
%\begin{widetext}
\begin{eqnarray} 
T_{1} : & \quad\left< 10^{+}_{1} \right>, \quad \left< 10^{-}_{1}
\right>\nonumber\\ 
T_{2}, T_{3}, T_{4} : & \quad \left< 10^{+}_{2,3,4} \right>
\nonumber\\ 
T_{5} : & \quad \left< 10^{-}_{5} \right>
\nonumber\\
\overline{C} : & \quad \left< \overline{126}^{-} \right>
\nonumber\\
\overline{C}_{1}, \overline{C}_{2} : & \quad \left< \overline{126}'^{0}_{1,2}
\right>
\end{eqnarray}
%\end{widetext}
and 
%\begin{widetext}
\begin{eqnarray}
\left< 10_{1}^{+} \right> = \left< 10_{3}^{+} \right>, \qquad
\left< 10_{1}^{-} \right> = \left< 10_{5}^{-} \right>\nonumber\\
\left< \overline{126}_{1}^{'0} \right> = \left< \overline{126}_{2}^{'0} \right>
\end{eqnarray}
%\end{widetext}
(Note that, with a $\overline{126}_{H}$ acquiring VEV, there must be a
conjugate $126_{H}$ acquiring VEV  to cancel the D-term. Since $126_{H}$ does
not couple to $16_{i}$, it has no role in the construction of the Yukawa
sector.) The needed flavon VEVs are given by 
%\begin{widetext}
\begin{eqnarray}
\label{eq:flavonvev}
\left< \phi_{(1)} \right> = 
\left( \begin{array}{c} \epsilon' \\ 0 \end{array} \right),
\qquad
\left< \phi_{(2)} \right> = 
\left( \begin{array}{c} 0 \\ \epsilon \end{array} \right)
\nonumber\\  
\left< S_{(1)} \right> = 
\left( \begin{array}{cc} 0 & \epsilon' \\ \epsilon' & 0 \end{array} \right),
\qquad
\left< S_{(2)} \right> = 
\left( \begin{array}{cc} 0 & 0  \\ 0 & \epsilon \end{array} \right)
\nonumber\\
\Phi = \left( \begin{array}{c} \delta_{1} \\ \delta_{3} \end{array} \right),
\qquad
\Sigma = \left( \begin{array}{cc} 0 & 0 \\ 0 & \delta_{2} \end{array} \right)
\end{eqnarray}
%\end{widetext}
Our $(Z_{2})^{3}$ charge assignments give rise to a unique superpotential: 
%\begin{widetext}
\begin{equation}
W = W_{Dirac} + W_{\nu_{RR}}
\end{equation}
\begin{eqnarray}
W_{Dirac}=\psi_{3}\psi_{3} T_{1}
 + \frac{1}{M} \psi_{3} \psi_{a}
\left(T_{2}\phi_{(1)}+T_{3}\phi_{(2)}\right)
\nonumber\\
+ \frac{1}{M} \psi_{a} \psi_{b} \left(T_{4} + \overline{C}\right) S_{(2)}
+ \frac{1}{M} \psi_{a} \psi_{b} T_{5} S_{(1)}
\nonumber\\
W_{\nu_{RR}}=\psi_{3} \psi_{3} \overline{C}_{1} 
+ \frac{1}{M} \psi_{3} \psi_{a} \Phi \overline{C}_{2}
+ \frac{1}{M} \psi_{a} \psi_{b} \Sigma \overline{C}_{2}
\end{eqnarray}
%\end{widetext} 
The mass matrices then can be read from the superpotential to be
%\begin{widetext}
%\rule{8.4cm}{.3pt}%
%\rule{.3pt}{.25cm}  
\begin{eqnarray}
M_{u,\nu_{LR}} & = &
\left( \begin{array}{ccc}
0 & 0 & \left<10_{2}^{+} \right> \epsilon'\\
0 & \left<10_{4}^{+} \right> \epsilon & \left<10_{3}^{+} \right> \epsilon \\
\left<10_{2}^{+} \right> \epsilon' & \left<10_{3}^{+} \right> \epsilon &
\left<10_{1}^{+} \right>
\end{array} \right)
\nonumber\\
 & = & 
\left( \begin{array}{ccc}
0 & 0 & r_{2} \epsilon'\\
0 & r_{4} \epsilon & \epsilon \\
r_{2} \epsilon' & \epsilon & 1
\end{array} \right) M_{U}
\end{eqnarray}
\begin{eqnarray}
M_{d,e} & = & 
\left(\begin{array}{ccc}
0 & \left<10_{5}^{-} \right> \epsilon' & 0 \\
\left<10_{5}^{-} \right> \epsilon' &  (1,-3)\left<\overline{126}^{-} \right>
\epsilon & 0\\ 0 & 0 & \left<10_{1}^{-} \right>
\end{array} \right)
\nonumber\\
 & = & 
\left(\begin{array}{ccc}
0 & \epsilon' & 0 \\
\epsilon' &  (1,-3) p \epsilon & 0\\
0 & 0 & 1
\end{array} \right) M_{D}
\end{eqnarray}
%\end{widetext}
where
%\begin{widetext}
\begin{equation}
\label{eq:higgsvev}
M_{U} \equiv \left<10_{1}^{+} \right>, 
\qquad
M_{D} \equiv \left<10_{1}^{-} \right>
\end{equation}
\begin{eqnarray}
r_{2} \equiv \left<10_{2}^{+} \right> / \left<10_{1}^{+} \right>,
\quad
& r_{4} \equiv \left<10_{4}^{+} \right> / \left<10_{1}^{+} \right>
\nonumber\\
p \equiv \left<\overline{126}^{-} \right> / \left<10_{1}^{-} \right>
&\end{eqnarray}
%\end{widetext}
The right-handed neutrino mass matrix is
%\begin{widetext}
%\rule{8.4cm}{.3pt}%
%\rule{.3pt}{.25cm}  
\begin{eqnarray}
M_{\nu_{RR}} & = &  
\left( \begin{array}{ccc}
0 & 0 & \left<\overline{126}_{2}^{'0} \right> \delta_{1}\\
0 & \left<\overline{126}_{2}^{'0} \right> \delta_{2} 
& \left<\overline{126}_{2}^{'0} \right> \delta_{3} \\ 
\left<\overline{126}_{2}^{'0} \right> \delta_{1}
& \left<\overline{126}_{2}^{'0} \right> \delta_{3} &
\left<\overline{126}_{1}^{'0} \right> \end{array} \right)
\nonumber\\
 & = & 
\left( \begin{array}{ccc}
0 & 0 & \delta_{1}\\
0 & \delta_{2} & \delta_{3} \\ 
\delta_{1} & \delta_{3} & 1
\end{array} \right) M_{R}
\end{eqnarray}
%\end{widetext}
with $M_{R} \equiv \left<\overline{126}^{'0}_{1}\right>$.
We have thus arrived at the mass matrices shown in eq.(\ref{eq:texture}),
(\ref{eq:Mnu}) and (\ref{eq:Mrr}).

%\section{RGE analysis}
%\label{sec:rge}
%\subsection{}
%\subsubsection{}
{\bf RGE Analysis and Results:}
In order to obtain the input parameters at the GUT scale, first
we need to know various Yukawa couplings (the diagonal elements) and mixing
angles at the GUT scale. We use the expressions derived from 1-loop RGEs given
by \cite{Arason:rge1992a,Berezhiani:fmass1999a}: 
%\begin{widetext}
\begin{eqnarray}
m_{u}=Y_{u}^{0}R_{u}\eta_{u}B_{t}^{3}v_{u},& \quad
m_{c}=Y_{c}^{0}R_{u}\eta_{c}B_{t}^{3}v_{u}\nonumber\\
m_{t}=Y_{c}^{0}R_{u}B_{t}^{6}v_{u} &\nonumber\\
m_{d}=Y_{d}^{0}R_{d}\eta_{d}v_{d},& \quad
m_{s}=Y_{s}^{0}R_{d}\eta_{s}v_{d}\nonumber\\
m_{b}=Y_{b}^{0}R_{d}\eta_{b}B_{t}v_{d} &\nonumber\\
m_{e}=Y_{e}^{0}R_{e}v_{d},& \quad
m_{\mu}=Y_{\mu}^{0}R_{e}v_{d}\nonumber\\
m_{\tau}=Y_{\tau}^{0}R_{e}v_{d} &
\end{eqnarray}
\begin{equation}
V_{ij} = \{
\begin{array}{lll}
V_{ij}^{0}, & & ij = ud, us, cd, cs, tb \\
V_{ij}^{0} B_{t}^{-1}, & & ij = ub, cb, td, ts.
\end{array}
\end{equation}
%\end{widetext}
where $V_{ij}$ are CKM matrix elements; quantities with superscript $0$ are
evaluated at GUT scale, and all the $m_{f}$ and $V_{ij}$ are the experimental
values \cite{PDG:exp1998a}. We will assume $\tan \beta =
\frac{v_{u}}{v_{d}}=10$ and 
$v=\sqrt{v_{u}^{2}+v_{d}^{2}}=\frac{246}{\sqrt{2}} GeV.$
The running factor $\eta_{f}$ includes QCD + QED contributions:  
For $f=b,c$, $\eta_{f}$ is for the range $m_{f}$ to $m_{t}$, and for
$f=u,d,s$, $\eta_{f}$ is for the range $1 GeV$ to $m_{t}$;
%\begin{widetext}
\begin{eqnarray*}
\eta_{u}=\eta_{d}=\eta_{s}=2.38_{-0.19}^{+0.24}\\
\eta_{c}=2.05_{-0.11}^{+0.13}\\
\eta_{b}=1.53_{-0.04}^{+0.03}.
\end{eqnarray*}
%\end{widetext}
$R_{u,d,e}$ are contributions of the gauge-coupling constants
running from weak scale $M_{z}$ to the SUSY breaking scale, taken to be
$m_{t}$, with the SM spectrum, and from $m_{t}$ to the GUT scale with MSSM
spectrum;
\begin{displaymath} 
R_{u}=3.53_{-0.07}^{+0.06}, \qquad 
R_{d}=3.43_{-0.06}^{+0.07}, \qquad 
R_{e}=1.50.
\end{displaymath}
$B_{t}$ is the running induced by large top-quark Yukawa coupling defined
by
%\begin{widetext}
\begin{equation}
B_{t}=\exp\left[\frac{-1}{16 \pi^{2}} \int_{\ln M_{SUSY}}^{\ln M_{GUT}}
Y_{t}^{2}(\mu) d(\ln \mu)\right]  
\end{equation}
%\end{widetext}
which varies from $0.7$ to $0.9$ corresponding to the perturbative
limit $Y_{t}^{0}\approx 3$ and the lower limit $Y_{t}^{0} \approx 0.5$ imposed
by the top-pole mass. 

In order to have a good fit to these values, we first obtain the following
approximate analytic expressions 
%\begin{widetext}
\begin{eqnarray} 
a \simeq \sqrt{
\frac{Y_{u}^{0}Y_{c}^{0}}{Y_{t}^{0}(Y_{c}^{0}+c^{2}Y_{t}^{0})}}, \quad & 
b \simeq c^{2} + \frac{Y_{c}^{0}}{Y_{t}^{0}}\nonumber\\
d \simeq Y_{t}^{0} & \nonumber\\
e \simeq \sqrt{\frac{Y_{e}^{0}}{Y_{\mu}^{0}-2Y_{e}^{0}}}
\frac{Y_{\mu}^{0}-Y_{e}^{0}}{Y_{\tau}^{0}}, \quad & 
f \simeq \frac{Y_{\mu}^{0}-Y_{e}^{0}}{3Y_{\tau}^{0}}\nonumber\\ 
h = Y_{\tau}^{0} & 
\end{eqnarray}
%\end{widetext}
With the GUT scale values of $Y_{e}^{0}, Y_{\mu}^{0}$ and $Y_{\tau}^{0}$, the
three parameters $e, f$, and $h$ in the down-quark and charged lepton sectors
are uniquely determined. 
With the GUT scale values of $Y_{u}^{0}, Y_{c}^{0}$ and $Y_{t}^{0}$,
these relations reduce the number of parameters in the up-quark and Dirac
neutrino sectors from four down to one, the parameter $c$. Using the
GUT scale value of the Cabbibo angle, $V_{us}$, the value of $c$ is
determined. 
At the GUT scale which is taken to be $M_{GUT} = 2.39 \times 10^{16} GeV$, 
with $g_{1} = g_{2} = g_{3} = 0.7530$,
our input parameters are chosen to be:  
%\begin{widetext}
\begin{eqnarray}
a = \alpha = 0.00226, & \quad b = \beta = 0.00381\nonumber\\ 
c = \gamma = 0.0328, & \quad d = \eta = 0.572\nonumber\\ 
e = 0.00403, & \quad f = 0.0195 \nonumber\\
h = 0.0678 &\nonumber\\ 
\delta_{1} = 0.00116, & \quad \delta_{2} = 3.32 \times 10^{-5}\nonumber\\
\delta_{3} = 0.0152 & \nonumber\\
M_{R} = 1.32 \times 10^{14} GeV & 
\end{eqnarray}
%\end{widetext}
These parameters are related to $\epsilon$ and $\epsilon'$ given in
eq.(\ref{eq:flavonvev}) and Higgs VEVs and their ratios given in
eq.(\ref{eq:higgsvev}) \cite{chen:fmass2000a}. 
$\delta_{i}$'s could be obtained using $t=1 \times 10^{-3}$ in eq.(\ref{eq:delta});
$\Lambda$ is expressible in terms of $\frac{\eta^{2}}{M_{R}}$ due to eq.(\ref{eq:seesaw}).
The Yukawa couplings in the down-quark and
charged lepton sectors are then given by
%\begin{widetext}
\begin{eqnarray}
Y_{d}^{0}= 0.00005441, \quad &
Y_{s}^{0}= 0.001374 \nonumber\\
Y_{b}^{0}= 0.06779  & \nonumber\\
Y_{e}^{0}= 0.00001880, \quad &
Y_{\mu}^{0}= 0.003979 \nonumber\\
Y_{\tau}^{0}= 0.06779 & 
\end{eqnarray}
%\end{widetext}
and various ratios are given by
%\begin{widetext}
\begin{equation}
\frac{Y_{d}^{0}}{Y_{e}^{0}}=2.895, \quad
\frac{Y_{s}^{0}}{Y_{\mu}^{0}}=\frac{1}{2.895}, \quad
\frac{Y_{b}^{0}}{Y_{\tau}^{0}}=1
\end{equation}
%\end{widetext}
which agree with Georgi-Jarlskog relations.

Having determined the GUT scale values of these elements, we then
numerically solve the one-loop RGEs for the MSSM spectrum with three
right-handed neutrinos \cite{Babu:rge1993a}   
from GUT scale to the effective right-handed neutrino mass scale, 
$M_{R} \simeq 1.32 \times 10^{14} GeV$.
At $M_{R}$, seesaw mechanism is implemented. We then run the MSSM RGEs
 \cite{Arason:rge1992a}
from $M_{R}$ down to the SUSY breaking scale $m_{t} \simeq 176 GeV$, and then
the SM RGEs from $m_{t}$ to $M_{z}=91.187 GeV$. The light neutrino RGEs \cite{Babu:rge1993a} 
are also used from $M_{R}$ to $M_{z}$. Predictions obtained at $M_{z}$ are
summarized in Table \ref{predict}, taking into account the SUSY threshold
corrections  \cite{Hall:rge1994a}
\begin{displaymath}
\Delta_{s} = -0.10, \qquad \Delta_{b} = -0.25
\end{displaymath}
They are to be compared with the values at $M_{z}$ calculated from the
experimental values by the authors of \cite{Fusaoka:exp1998a}.
%\pagebreak  
%\begin{widetext}
\begin{table}
%\begin{widetext}
\begin{tabular}{l c | c c l c c c l c c c l}
\hline
 & & & & data at $M_{z}$ \qquad \qquad
 & & & & predictions \qquad 
 & & & & predictions  \\ 
& & & & 
 & & & & at $M_{GUT}$ \qquad 
 & & & & at $M_{z}$\\ 
\hline
$m_{u}$
& & & & $2.33^{+0.42}_{-0.45}MeV$ 
& & & & $4.065 \times 10^{-6}$
& & & & $1.917 MeV$\\
$m_{c}$ 
 & & & & $677^{+56}_{-61}MeV$  
 & & & & $0.001566$
 & & & & $738.7 MeV$\\
$m_{t}$ 
 & & & & $181^{+}_{-}13GeV$ 
 & & & & $0.5729$
 & & & & $184.3 MeV$\\
$\frac{m_{d}}{m_{s}}$  
& & & & $17 \sim 25$  
& & & & $3.96 \times 10^{-2}$
& & & & $22.5$\\
$m_{s}$ 
& & & & $93.4^{+11.8}_{-13.0}MeV$  
& & & & $0.001374$
& & & & $83.15 GeV$\\
$m_{b}$ 
& & & & $3.00^{+}_{-}0.11GeV$  
& & & & $0.06779$
& & & & $3.0141 GeV$\\
$m_{e}$  
& & & & $0.486847MeV$  
& & & & $1.880 \times 10^{-5}$
& & & & $0.486 MeV$\\
$m_{\mu}$  
& & & & $102.75MeV$ 
& & & & $0.003979$ 
& & & & $102.8 MeV$\\
$m_{\tau}$  
& & & & $1.7467 GeV$  
& & & & $0.06779$
& & & & $1.744 GeV$ \\
\hline
\end{tabular}
%\end{widetext} 
\caption{\linespread{0.9} \scriptsize Predictions and values extrapolated
from experimental data. The first column shows the results calculated by
Fusaoka et al \cite{Fusaoka:exp1998a}. The second column shows the Yukawa
couplings at the GUT scale obtained with the input parameters we have chosen.
The third column shows the predictions at $M_{z}$ after renormalization group
effects have been taken into account.}    
\label{predict} 
\end{table}
%\end{widetext}
The quark mixing matrix $V_{CKM}$ at $M_{z}$ is predicted to be 
\begin{eqnarray}
\left|V_{CKM, predict}\right| =
|V_{u_{L}} V_{d_{L}}^{\dag}|
\nonumber\\
=\left(\begin{array}{ccc}
0.9751 & 0.2215 & 0.003541 \\
0.2215 & 0.9745 & 0.03695 \\
0.004735 & 0.03681 & 0.9993
\end{array} \right)
\end{eqnarray}
They are to be compared with the experimental results
extrapolated to $M_{z}$ \cite{Fusaoka:exp1998a} 
\pagebreak
\begin{widetext}
%\noindent
%\rule{8.8cm}{.3pt}%
%\rule{.3pt}{.25cm}  
\begin{equation}
\left|V_{CKM, exp}\right|=
\left(\begin{array}{ccc}
0.9745-0.9757 & 0.219-0.224 & 0.002-0.005 \\
0.218-0.224 & 0.9736-0.9750 & 0.036-0.046 \\
0.004-0.014 & 0.034-0.046 & 0.9989-0.9993
\end{array} \right)
\end{equation}
\end{widetext}

Our model predicts the three light Majorana neutrino masses to be
%\begin{widetext}
\begin{eqnarray}
m_{\nu_{1}} = 2.0052 \times 10^{-4} eV \nonumber\\
m_{\nu_{2}} = 2.0123 \times 10^{-4} eV \nonumber\\
m_{\nu_{3}}= 0.05574 eV
\end{eqnarray}
%\end{widetext}
and the resulting squared mass differences are
\begin{eqnarray}
\Delta m_{23}^{2} = 3.11 \times 10^{-3} eV^{2} \nonumber\\
\Delta m_{12}^{2} = 2.87 \times 10^{-10} eV^{2}
\end{eqnarray}
The lepton mixing matrix is given by
%\begin{widetext}
\begin{eqnarray}
\left|U_{MNS, predict}\right| =
\left|U_{e_{L}} U_{\nu_{LL}}^{\dag} \right| \nonumber\\
=
\left(\begin{array}{ccc}
0.6710 & 0.7396 & 0.0527 \\
0.5410 & 0.4397 & 0.7169 \\
0.5070 & 0.5096 & 0.6952
\end{array} \right)
\end{eqnarray}
%\end{widetext}
This translates into
%\begin{widetext}
\begin{eqnarray}
\sin^{2} 2\theta_{atm} \equiv
4|U_{\mu\nu_{3}}|^{2}(1-|U_{\mu\nu_{3}}|^{2}) =0.9992 \nonumber\\
\sin^{2} 2\theta_{\odot} \equiv 4|U_{e\nu_{2}}|^{2}(1-|U_{e\nu_{2}}|^{2})
= 0.9912.
\end{eqnarray}
%\end{widetext} 
These values agree with the Super-Kamiokande
atmospheric neutrino oscillation data \cite{SuperK:1998a,LP99atm}, and the
solar VO solution \cite{LP99solar}.  And the $(1,3)$ element of $U_{MNS}$ is
given by $|U_{e \nu_{3}}| = 0.0527$ which is far below the bound by the
CHOOZ experiment  $|U_{e\nu_{3}}| \lesssim 0.16$ \cite{Apollonio:1999ae}. 
The three eigenvalues of the right-handed neutrino Majorana mass matrix are
given by
%\begin{widetext}  
\begin{eqnarray}
M_{RR_{1}} \simeq 2.963 \times 10^{7} GeV \nonumber\\
M_{RR_{2}} \simeq 2.643 \times 10^{10} GeV \nonumber\\
M_{RR{3}} \simeq 1.319 \times 10^{14} GeV
\end{eqnarray}
%\end{widetext}

We can have the LOW solution (-- a LAMSW solution with $\Delta m_{12}^{2}
\sim 10^{-6} - 10^{-7} eV^{2}$) with
\begin{equation}
\delta_{1}= 0.001147, \quad
\delta_{2}= 0.0002354, \quad
\delta_{3}= 0.01675
\end{equation}
\begin{displaymath}
M_{R}= 1.615 \times 10^{13} GeV
\end{displaymath}
These change the predictions of $m_{u,c,t}$ by less than $1\%$ but have no
observable effects on down-quark and charged lepton masses, and
the CKM matrix remains essentially the same \cite{chen:fmass2000a}. In
the neutrino sector, we get 
\begin{eqnarray}
m_{\nu_{1}}=0.001626 eV \nonumber\\
m_{\nu_{2}}=0.001650 eV \nonumber\\
m_{\nu_{3}}=0.06303 eV 
\end{eqnarray}
and the squared mass differences are
%\begin{widetext}
\begin{eqnarray}
\Delta m_{23}^{2} = 3.973 \times 10^{-3} eV^{2} \nonumber\\
\Delta m_{12}^{2} = 1.298 \times 10^{-7} eV^{2}
\end{eqnarray}
%\end{widetext}
The lepton mixing matrix is given by
\begin{eqnarray}
\left|U_{MNS, predict}\right| = 
|U_{e_{L}} U_{\nu_{LL}}^{\dag}| \nonumber\\
= \left(\begin{array}{ccc}
0.6665 & 0.7418 & 0.07428 \\
0.5511 & 0.4231 & 0.7192 \\
0.5021 & 0.5202 & 0.6909
\end{array} \right)
\end{eqnarray}
The element $| U_{e \nu_{3}} |$ is predicted to be 0.07428, 
which is less than the experimental upper bound.
The three right-handed neutrino eigenvalues are predicted to be 
\begin{eqnarray}
M_{RR_{1}} \simeq 9.558 \times 10^{7} GeV \nonumber\\
M_{RR_{2}} \simeq 8.453 \times 10^{8} GeV \nonumber\\
M_{RR_{3}} \simeq 1.615 \times 10^{13} GeV
\end{eqnarray}

It is also possible to have the LAMSW solution with
\begin{equation}
\delta_{1}= 0.001082, \quad
\delta_{2}= 0.0009870, \quad
\delta_{3}= 0.02238
\end{equation}
\begin{displaymath}
M_{R}= 2.415 \times 10^{12} GeV
\end{displaymath}
The predictions in the quark and the charged lepton sectors remain the same. 
In the neutrino sector, we get 
\begin{eqnarray}
m_{\nu_{1}}=0.01089 eV \nonumber\\
m_{\nu_{2}}=0.01206 eV \nonumber\\
m_{\nu_{3}}=0.09999 eV 
\end{eqnarray}
and the squared mass differences are
%\begin{widetext}
\begin{eqnarray}
\Delta m_{23}^{2} = 9.851 \times 10^{-3} eV^{2} \nonumber\\
\Delta m_{12}^{2} = 2.752 \times 10^{-5} eV^{2}
\end{eqnarray}
%\end{widetext}
The lepton mixing matrix is given by
\begin{eqnarray}
\left|U_{MNS, predict}\right| = 
|U_{e_{L}} U_{\nu_{LL}}^{\dag}| \nonumber\\
= \left(\begin{array}{ccc}
0.6439 & 0.7486 & 0.1580 \\
0.6045 & 0.3712 & 0.7049 \\
0.4690 & 0.5494 & 0.6915
\end{array} \right)
\end{eqnarray}
The element $| U_{e \nu_{3}} |$ is predicted to be 0.1580 which is right at
the experimental bound $| U_{e \nu_{3}} | \lesssim 0.16$
\cite{Apollonio:1999ae}.  The three right-handed neutrino eigenvalues
are given by 
\begin{eqnarray}
M_{RR_{1}} \simeq 5.732 \times 10^{6} GeV \nonumber\\
M_{RR_{2}} \simeq 1.177 \times 10^{9} GeV \nonumber\\
M_{RR_{3}} \simeq 2.417 \times 10^{12} GeV
\end{eqnarray}
We note that a $|U_{e\nu_{3}}|$ value of less than 0.1580 would lead to
$\Delta m_{23}^{2} > 10^{-2} eV^{2}$  
leading to the elimination of the LAMSW
solution in our model. This is a characteristic of the LAMSW solution with 
$\Delta m_{12}^{2} \gtrsim 10^{-5} eV^{2}$.

\textit{Note added:}
The form invariance of eq.(\ref{eq:seesaw}) -- $M_{\nu_{LL}}$ having the same
texture as that of $M_{\nu_{LR}}$ and $M_{\nu_{RR}}$ -- also occurs in a model
of neutrino mixing \cite{Fritzsch:fmass1999a} which uses different symmetric
mass textures.

{\bf Summary:}
We have constructed a realistic model based on SUSY $SO(10)$ combined with
$U(2)$ flavor symmetry. The up-quark sector is related to the Dirac neutrino
sector, and the down-quark sector is related to the charged lepton sector via
$SO(10)$ symmetry. The inter-family hierarchy is achieved via $U(2)$ symmetry.
In contrast to the commonly used effective operator approach, we use $126$-dim
Higgses to construct the Yukawa sector. R-parity symmetry is thus
automatically preserved at low energies. In our model, the Dirac and 
right-handed Majorana neutrino mass matrices which have very small mixing
combine with the seesaw mechanism resulting in a large mixing in the lepton
sector. The symmetric mass textures arising from the left-right symmetry
breaking chain of $SO(10)$ which we have considered give rise to very good predictions; 
15 masses (including the right-handed neutrino masses) and 6 mixing angles are
predicted by 11 parameters. Our model favors the vacuum oscillation 
and LOW solutions to the solar neutrino problem.

% If in twocolumn mode, this environment will move to single column
% format so that long equations can be displayed. Use
% Note: this may cause bad behavior if this occurs near a page
% break - this can be worked around by adding an explicit \pagebreak.
%\begin{widetext}
%\end{widetext}

\pagebreak

% If you have acknowledgments this puts in the proper section
\begin{acknowledgments}
We thank K.S. Babu, C. Carone, N. Irges and S. Oh for useful communications. 
This work was supported, in part, by the US Department of Energy Grant No. DE
FG03-05ER40894.  
\end{acknowledgments}

%\pagebreak
% Create the reference section using BibTeX
\bibliography
{BIBFILE/fmass,BIBFILE/dts,BIBFILE/pdecay,BIBFILE/rge,BIBFILE/exp,BIBFILE/group,BIBFILE/rparity,BIBFILE/pheno,BIBFILE/so10}

% figures follow here or may be put into the text as floats.
% Use the graphics or
% graphicx packages distributed with LaTeX2e. See the LaTeX Graphics
% Companion by Michel Goosens, Sebastian Rahtz, and Frank Mittelbach
% for instance.
% \begin{figure}
% \label{}
% \includegraphics[]{}
% \caption{}
% \end{figure}

% tables follow here or maybe be put in the text
% \begin{table}
% \label{}
% \caption{}
% \end{tabular}
% \end{table}

%\appendix

\end{document}